\documentclass{aa}
\usepackage{txfonts}
\usepackage{graphicx}
\def\msun{\hbox{M$_\odot$}}

\begin{document}

\title{APOGEE physical properties of globular cluster tidal tails}

\author{A.E. Piatti\inst{1,2,}\thanks{\email{andres.piatti@fcen.uncu.edu.ar}}}

\institute{Instituto Interdisciplinario de Ciencias B\'asicas (ICB), CONICET-UNCuyo, Padre J. Contreras 1300, M5502JMA, Mendoza, Argentina;
\and Consejo Nacional de Investigaciones Cient\'{\i}ficas y T\'ecnicas (CONICET), Godoy Cruz 2290, C1425FQB,  Buenos Aires, Argentina\\
}

\date{Received / Accepted}

\abstract{A recent model prediction claimed that exists a correlation between the formation 
scenarios of globular clusters, i.e., whether they formed in situ, or in dark matter halos that
were accreted into the Milky way, with some properties of their tidal tails, particularly,
their widths ($w$), their dispersion in the z-component of the angular momentum 
($\sigma$$_{\rm L_z}$ ), and in the line-of-sight ($\sigma$$_{\rm V_{LOS}}$) and 
tangential ($\sigma$$_{\rm V_{Tan}}$) velocities. I exploited the APOGEE DR17 data base 
and selected highly confident tidal tails members of 17 Milky Way globular clusters, for which 
the above four properties were computed for the first time. From all possible paired
property combinations, I found that $\sigma$$_{\rm V_{LOS}}$ and $\sigma$$_{\rm V_{Tan}}$ 
resulted to be highly correlated, nearly to the identity relationship. This 
observation-based correlation resulted to be in an overall very good agreement with that 
arising from the aforementioned predictions. Additionally, when the four analyzed properties
are linked to the accretion groups of the Milky way to which the globular clusters
are meant to be associated, I found kinematically cold and hot tidal tails pertaining to
globular clusters dsitributed in all the considered accretion groups. This outcome could be
an evidence that globular clusters form in galaxies within a wide variety of dark matter
halos, with different masses and profiles.}


\titlerunning{APOGEE tidal tails}  

\authorrunning{A.E. Piatti and K. Malhan}           

\maketitle

\markboth{A.E. Piatti and K. Malhan:}{APOGEE tidal tails}

\section{Introduction}           

Recently, \citet{malhanetal2021} and \citet{malhanetal2022} showed that 
there is a correspondence between some physical properties of globular 
cluster tidal tails and the formation environment of those globular 
clusters, namely, whether they were accreted or formed in situ. Particularly,
they found that the width of the tidal tails, and the dispersion in the 
z-component of the angular momentum, and in the line-of-sight and tangential 
velocities of member stars of globular cluster tidal tails tell us about the 
globular clusters’ origin. They simulated globular clusters formed in a low-mass 
galaxy halo with cored or cuspy central density profiles of dark matter that 
later merged with the Milky Way and globular clusters formed in situ. On average, 
globular clusters from cuspy profiles have tidal tails with  dispersion in the 
z-component of the angular momentum, and in the line-of-sight and tangential velocities
three times larger than those of clusters formed in cored dark matter
profiles. Likewise, globular clusters formed in situ have mean values of
dispersion of these properties nearly ten times smaller than those for 
globular clusters accreted inside cored subhalos. As for the tidal tails' width,
it increases from some tens of parsecs for those formed in situ to few hundred
parsecs for globular clusters formed inside cored dark matter profiles, to nearly 
a couple of kilo parsecs for cuspy ones.

These predictions open, for the first time, the possibility of disentangling 
the origin of individual globular clusters without the need of constructing a 
multicomponent model for the whole population of galactic globular clusters 
\citep[see, e.g.][]{massarietal2019,forbes2020,callinghametal2022}. However,
the usefulness of the predictions depends on the knowledge of the
aforementioned quantities for a statistically significant sample of member
stars of globular cluster tidal tails. As far as I are aware, out of all
Milky Way globular clusters with tidal tails \citep{pcb2020,zhangetal2022}, 
NGC~288 and NGC~~5904 (M5) are the only globular clusters with a tangential velocity 
dispersion analysis. \citet{piatti2023b} used 50 highest ranked tidal tail 
member candidates in M5 \citep{g2019} to conclude that the globular cluster 
formed inside a $\sim$ 10$^9$ $\msun$ cuspy dark matter subhalo. 
Likewise, \citet{grillmair2025} found that NGC~288's tidal tails are consistent
with the cluster being stripped from a parent galaxy with a large, cored
dark matter halo, which is in very good agreement with the results by 
\citet{belokurovetal2018} and \citet{helmietal2018}, who found that it could have 
been brought into the Milky Way halo during the Gaia-Enceladus-Sausage event. 

The Apache Point Observatory Galactic Evolution Experiment 
\citep[APOGEE][]{majewskietal2017} bring us the opportunity 
to exploit an homogeneous chemical compositions and kinematics data base for a 
large percentage of the Milky Way globular cluster population, with the aim of 
computing the width of tidal tails, and the dispersion of the z-component of the 
angular momentum, and in the 
line-of-sight and tangential velocities of globular cluster tidal tail members.
Precisely, the main aim of this work is to probe the performance of the
predictions by \citet{malhanetal2021} and \citet{malhanetal2022} by
comparing their outcomes with the observation-based dispersion values of globular 
clusters with data available in APOGEE. In Section~2 I describe the
employed data and the selection of globular cluster tidal tail members, and
 the computation of the dispersion values of the four
mentioned astrophysical properties. I analyze and discuss our findings in Section~3, and
in Section~4 I summarize the main conclusions of this work.

\section{Data handling}

\citet{schiavonetal2024} compiled a catalog for 72 Milky Way globular clusters
with information of the chemistry (20 species), radial velocities and imported
{\it Gaia} EDR3 data \citep{gaiaetal2021} of 6422 member stars, retrieved from 
APOGEE DR17 \citep{abdurroufetal2022}. The stars contained in the \citet{schiavonetal2024}'s 
APOGEE value-added catalog are mainly red giants for which the most reliable
abundances correspond to C, N, O, Mg, Al, Si, Mn, Fe,  and Ni. They defined
two types of cluster members, namely: likely members and outliers; the
former being those with the highest probabilities of being cluster members. 
As a stringent selection criterion, I used here only the likely members, and the
abundances of the eight most reliable chemical elements mentioned above, besides
[Fe/H].

With the aim of extracting from APOGEE DR17 tidal tails stars of Milky
Way globular clusters, I rely on their abundances of chemical elements, which
have negligible variations along their lifetime. Until recently, the mean proper
motion and/or radial velocity of globular clusters were used as selection
criteria of tidal tail stars, assuming that the tidal tail stars should share the
motions of cluster's stars \citep{sollima2020,xuetal2024}. However, 
since tidal tail stars escape from the cluster, they need to reach velocities
different than those of the cluster's members. In addition, the Milky Way
gravitational field accelerates them differently, so that the mean kinematic
properties of tidal tail stars vary along  the tail extensions 
\citep{piattietal2023,grondinetal2024}. Because of the diverse orientation in
space of globular cluster tidal tails, the heliocentric distances of their
stars vary along the tails, so that it would not seem appropriate to adopt a 
constant distance for them. This also prevent us against employing color-magnitude 
diagrams to select tidal tails stars placed along the cluster main sequences,
since the ordinate variable of the color-magnitude diagram depends on the star's 
heliocentric distance.

I first used the \citet{schiavonetal2024}'s APOGEE value-added catalog with the 
aim of computing the mean values and dispersion of [Fe/H] and those for the other 
eight chemical elements of the 72 Milky Way globular clusters.  Table~\ref{tab1a}
lists the resulting values. I did not include Ter~5, because it is a Milky Way
bulge globular cluster with a complex formation and evolution history and a
remarkable metallicity spread \citep{origliaetal2025}. For each globular cluster,
I then searched the APOGEE DR17 data base for stars with chemical element abundance 
patterns within 1$\sigma$ of that globular cluster, to be consistent with the
definition of likely members of \citet{schiavonetal2024} used here, which is a way to make 
a conservative selection leading up to a very pure sample. This
provides a high signal-to-noise ratio for contrasting the tidal tail profiles 
to the contamination. I retrieved only stars 
complying with the requisite:\\

[Fe/H]$_{\rm star}$ - $<$[Fe/H]$_{\rm cls}$$>$ $\le$ $\sigma$([Fe/H]$_{\rm cls}$)\\

and \\

\begin{figure}
\includegraphics[width=\columnwidth]{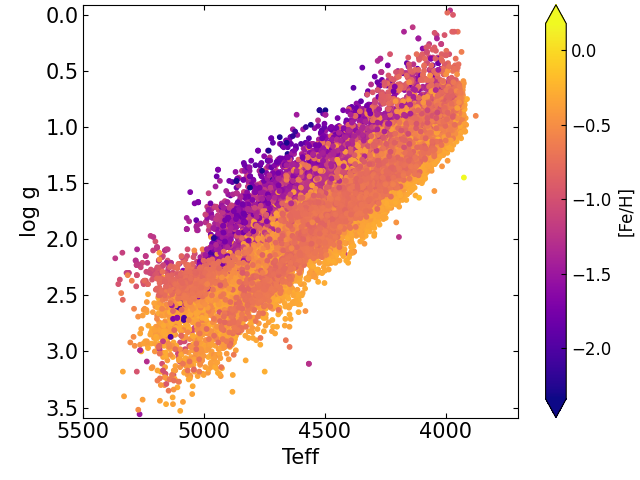}
\caption{log $g$ vs. $T_{\rm eff}$ plane with all the stars retrieved from the
APOGEE DR17 data base (see text for details).}
\label{fig1}
\end{figure}

[X/Fe]$_{\rm star}$ - $<$[X/Fe]$_{\rm cls}$$>$ $\le$ $\sigma$([X/Fe]$_{\rm cls}$)\\

\noindent where X represent the chemical elements C, N, O, Mg, Al, Si, Mn, and Ni,
and 'star' and 'cls' subscripts refer to the values of the searched stars
in the APOGEE DR17 data base and the clusters'  mean values and dispersion of 
Table~\ref{tab1a}. From this search, I found 42 and 12 globular clusters with no 
selected stars and few stars, respectively, that I removed from the final sample.
Figure~\ref{fig1} depicts the distribution of the 23917 finally extracted 
stars for the 17 surviving globular clusters in the log $g$ versus $T_{\rm eff}$ 
plane. As can be seen, they are all red giant stars.

The similarity imposed between the chemical element abundance patterns of tidal tail 
and cluster stars does not warrant that the selected stars are indeed globular cluster 
tidal tails ones. As is known, the Milky Way disks and halo are populated by stars
with metallicity patterns that are indistinguishable from those of globular
clusters \citep{reddyetal2003,vennetal2004,reddyetal2006,ns2010}. In order to remove
any potential star not belonging to globular cluster's tidal tails, I required
that the selected stars trace a smooth stellar path from the globular cluster's
center; are located beyond the globular cluster's radius, and show a smooth correlation
of their widths and of their z-component of the angular momentum,  line-of-sight and tangential 
velocities with the position along the tidal tails. I relied to this respect in the
recipe outlined by \citet{malhanetal2021} and \citet{malhanetal2022} to compute 
dispersion of different properties along tidal tails. Briefly, they defined an angular coordinates along 
the tidal tails and another perpendicular to them. Then they fitted the variation of each
considered property along 30$\degr$ long segments of tidal tails with a smooth polynomial, and
subtracted the fitted function from the properties values to obtain the residual
distribution. The standard deviation of this distribution provides the property dispersion
of that tidal tail segment. These criteria aimed at selecting a coherent 
sample of stars highly ranked as globular cluster tidal tails members.

For that purpose, 
I used the globular clusters's central coordinates (Right Ascension, Declination), 
mean heliocentric distances and radial velocities from \citet{schiavonetal2024}, and
{\it Gaia} astrometry information (parallax, proper motion) digested into the 
APOGEE DR17 data base \citep{abdurroufetal2022}, and followed the recipes 
summarized above \citep[see also,][]{piatti2025b}
to compute the variation in the four considered properties along the
physical extension of the tidal tails. In brief, I 
computed the Cartesian coordinates ($X$,$Y$,$Z$) and space velocity components
($V_X$, $V_Y$, $V_>$) of each selected star, and from them I obtained the z-component of their 
angular momentum. I also calculated their tangential velocities
V$_{\rm  Tan}$ = $k$ $\times$ (1/parallax) $\times$ $\mu$; where $k$ = 4.7405 km 
s$^{\rm -1}$ kpc$^{\rm -1}$ (mas/yr)$^{\rm -1}$ and $\mu$ is the total proper motion.
Then, I carried out two perpendicular rotations around the cluster's center in order to 
have the tidal tails centered on the cluster and mainly oriented along one of the three 
perpendicular axes ($\phi_1$). Firstly, I rotated the galactic ($X,Y,Z$) system 
around the $Y$ axis  to the ($X',Y,\phi_3$) system, as follows:\\

$X' = X cos(\theta) + Z sin(\theta)$\\

$Y$ = Y\\

$\phi_3 = -X sin(\theta) + Z cos(\theta)$,\\

\noindent where $\theta$ is the rotating angle to have the tidal tail in the ($X,Z$) plane
aligned along the $X'$ direction. Then, I rotated the ($X',Y,\phi_3$) system around
the $\phi_3$ axis to the ($\phi_1,\phi_2,\phi_3$) system, as follows:\\

$\phi_1 =  X' cos(\psi) - Y sin(\psi)$\\

$\phi_2 = X' sin(\psi) + Y cos(\psi)$\\

\begin{figure}
\includegraphics[width=\columnwidth]{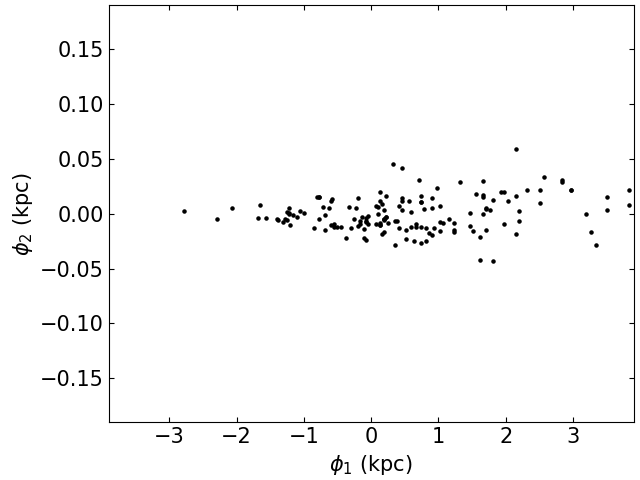}
\caption{Spatial diatribution in the ($\phi_1$,$\phi_2$) plane \citep[see][]{piatti2025b}
of selected tidal tails stars of NGC~5139.}
\label{fig2}
\end{figure}

$\phi_3 = \phi_3$\\,

\noindent where $\psi$ is the rotating angle to have the tidal tail in the ($X',Y$) plane
aligned along the $\phi_1$ direction. The appropriate rotation angles $\theta$ and $\psi$
were obtained by visually inspecting the orientation of the
stellar density contours in the rotated 3D coordinate system. I called the rotated framework 
($\phi$$_1$,$\phi$$_2$,$\phi$$_3$), where $\phi$$_1$ is along the tidal tails, $\phi$$_2$
is perpendicular to $\phi$$_1$ and contained in the tidal tails plane, and $\phi$$_3$ is
perpendicular to $\phi$$_1$ and $\phi$$_2$.  Figure~\ref{fig2} illustrates the projected spatial
distribution in the ($\phi_1$,$\phi_2$) plane of the selected tidal tails stars of 
NGC~5139.  Figure~\ref{fig2append} includes similar figures of the 16 remaining globular clusters,
while the source identifications of the selected stars are compiled in Table\ref{tab3a}. 
Reasons supporting the choice of Cartesian galactocentric coordinates instead 
of angular ones and the transformation equations of both perpendicular rotations are
described in \citet{piatti2025b}.

\begin{figure}
\includegraphics[width=\columnwidth]{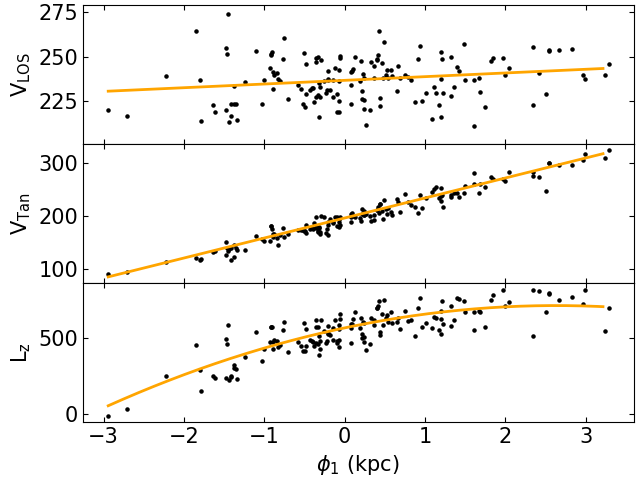}
\caption{V$_{\rm LOS}$ (km/s), V$_{\rm Tan}$ (km/s), and L$_{\rm z}$ (km/s kpc) as 
a function of $\phi_1$ for  selected tidal tails stars of NGC~5139 drawn with black 
dots. The orange line is the best-fitted polynomial.}
\label{fig3}
\end{figure}

I then plotted the $z$-component of the angular momentum (L$_{\rm z}$), and
the line-of-sight (V$_{\rm LOS}$) and tangential (V$_{\rm Tan}$) velocities as a function of 
$\phi_1$ for all the selected stars. Figure~\ref{fig3} illustrates the resulting spatial 
distributions of V$_{\rm LOS}$, V$_{\rm Tan}$, and L$_{\rm z}$ along the tidal tails direction 
($\phi_1$) of NGC~5139. The observed distributions were fitted with polynomials of up to 2nd order, 
which represent the mean behavior of these parameters along the tails. 
I visually inspected the fits of the properties of the 17 globular clusters, and they
readily visibly match the ensemble of points. Finally, I calculated 
the residuals (observed individual value - mean fitted value for the corresponding $\phi_1$)
and the respective standard dispersion, namely: $\sigma$$_{\rm V_{LOS}}$, $\sigma$$_{\rm V_{Tan}}$, 
and $\sigma$$_{\rm L_z}$, which are properties proposed by \citet{malhanetal2021} and 
\citet{malhanetal2022} sensible to the origin of Milky Way globular clusters, i.e., whether
formed in situ or in dwarf galaxies with central cored or cuspy dark matter profiles.
The tidal tails' width ($w$) was computed by calculating the average diameter of the
tidal tail's dispersion in the ($\phi_2$,$\phi_3$) plane. The uncertainties in these quantities 
are based on the errors in parallax, proper motions,
and radial velocity. I generated random kicks proportional to the error, which means 
that I sampled from the actual measurement from the uniform distribution within the interval
[measurement - error, measurement + error]. I randomly generated a thousand different values of parallax,
proper motions and radial velocity based on their
errors  (variable = mean\_value + variable\_error$\times$2$\times$(0.5-$t$), with -1$<$$t$$<$1 
randomly chosen), and
repeated the computation of all the quantities following the procedure described above. 
I used all the 1000 different $w$, 
$\sigma$$_{\rm V_{LOS}}$, $\sigma$$_{\rm V_{Tan}}$, and $\sigma$$_{\rm L_z}$ values to calculate 
their mean and dispersion. I checked that by using subsamples of 100 values did result in 
negligible mean and dispersion with respect to whole sample.
Table~\ref{tab2a} lists the resulting $w$, $\sigma$$_{\rm V_{LOS}}$, $\sigma$$_{\rm V_{Tan}}$, 
and $\sigma$$_{\rm L_z}$ values for 17 Milky Way globular clusters. Its second column
shows the length of the tidal tails, calculated as the difference in $\phi_1$ between the two
farthest stars located in opposite sides from the cluster's center, respectively. The resulting 
$\sigma$$_{\rm V_{Tan}}$ value of NGC~5904 is in very good agreement with that 
derived by \citet[7.50$\pm$1.38 km/s]{piatti2023b} using tangential velocities 
of 50 highly ranked tidal tails members.

\section{Analysis and discussion}

As far as I am aware, Table~\ref{tab2a} is the largest compilation of globular clusters'
tidal tails with their widths, and dispersion in the z-component of the angular momentum,
and in the line-of-sight and tangential velocities derived from observational data of
selected members. Therefore, it represents a valuable data set to explore whether
there exists any correlation between these properties, how well they match the models of
\citet{malhanetal2021} and \citet{malhanetal2022}, and what they tell us about the
association of the globular clusters to the different accretion groups of the Milky Way 
\citep[e.g.][]{callinghametal2022}. Furthermore, it can help not only to disentangling whether 
these parameters are drivers of the globular clusters' origin, but also to show the pace
of any eventual trends. For these reasons, I started by digging into the relationships of 
each one of these parameters with respect to the remaining three others.

\begin{figure*}
\includegraphics[width=\textwidth]{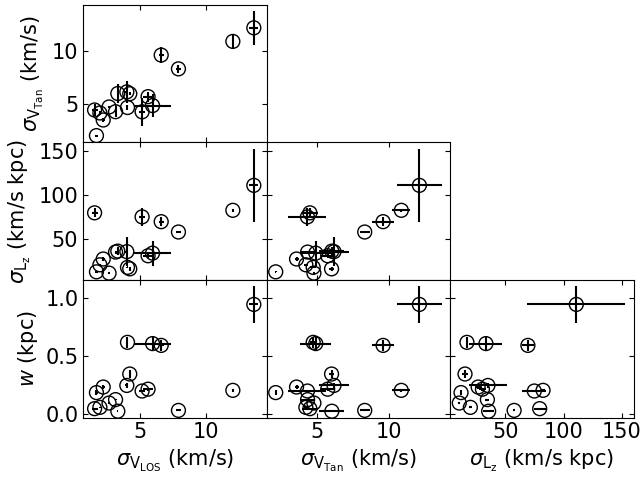}
\caption{Correlation between different globular cluster's tidal tails properties
using information of Table~\ref{tab2a}.}
\label{fig4}
\end{figure*}

Figure~\ref{fig4} depicts the collection of all the attempted relationships. 
At first glance, $\sigma$$_{\rm V_{LOS}}$ and $\sigma$$_{\rm V_{Tan}}$ seem to be 
tightly correlated;  $\sigma$$_{\rm L_z}$ shows a mild linear correlation with both
velocity dispersion, while $w$ does not seem to show any robust trend with
any of the other parameters. In order to quantify the visually suggested
behaviors, I have computed  bootstrap confidence intervals for two different
statistics, namely, Pearson and Spearman correlations, and the resulting
values are shown in Table~\ref{tab1}, which confirm the suggested trends. 
The confidence intervals were computed using \texttt{scipy.stats.bootstrap}
\footnote{https://docs.scipy.org/doc/scipy/reference/generated/scipy.stats.bootstrap.html}.
For the sake of the reader an example is provided below:

\includegraphics[width=\columnwidth]{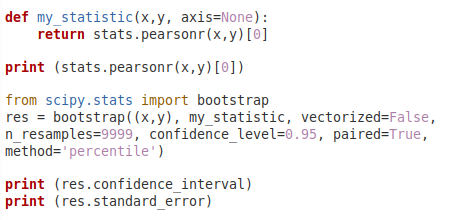}

These outcomes show that the physical width of tidal tails would not be necessarily
linked with their kinematic agitation, in the sense that the kinematically
hotter a tidal tail ($\sigma$ $>$ 1-2 km/s), the wider its extension, as expected.
Although some indication that the latter could be true, the tidal tails
sample analyzed here could be hampering such a trend. Nevertheless, the present
results also allow for the possibility that some hot tidal tails ($\sigma$ $>$ 5 km/s) 
are relatively thin ($w$ < 300 pc). 

\begin{table*}
\centering
\caption{ Bootstrap confidence intervals of the correlation between different globular cluster's tidal tails properties
using information of Table~\ref{tab2a}.}
\label{tab1}
\begin{tabular}{lcccccccc}
\hline\hline
Correlation       &  \multicolumn{4}{c}{Spearman} & \multicolumn{4}{c}{ Pearson} \\ 
                  &   statistic & low & high & std error &  statistic & low & high & std error \\

\hline
$\sigma$$_{\rm V_{LOS}}$ - $\sigma$$_{\rm V_{Tan}}$ &  0.80  & 0.45 & 0.94 & 0.13  &  0.91 & 0.68 & 0.97 & 0.07\\
$\sigma$$_{\rm L_z}$ - $\sigma$$_{\rm V_{LOS}}$     &  0.52  & 0.05 & 0.87 & 0.24  &  0.72 & 0.14 & 0.95 & 0.21\\
$\sigma$$_{\rm L_z}$ - $\sigma$$_{\rm V_{Tan}}$     &  0.55  & 0.03 & 0.90 & 0.22  &  0.72 & 0.22 & 0.95 & 0.19\\
$w$ - $\sigma$$_{\rm V_{LOS}}$                      &  0.48  & 0.02 & 0.83 & 0.22  &  0.55 & -0.02 & 0.90 & 0.25\\
$w$ - $\sigma$$_{\rm V_{Tan}}$                      &  0.28  & 0.26 & 0.68 & 0.24  &  0.47 & -0.20 & 0.84 & 0.28\\
$w$ - $\sigma$$_{\rm L_z}$                          &  0.01  & -0.52& 0.51 & 0.27  &  0.33 & -0.40 & 0.75 & 0.32\\\hline
\end{tabular}
\end{table*}

From a purely kinematic point of view, it would seem that the tidal tails
are similarly agitated in any observed direction, as judged by the nearly identity
relationship found between the dispersion of perpendicular velocities (V$_{\rm LOS}$ 
versus V$_{\rm Tan}$). Whenever a physical position parameter is combined with
the kinematic behavior, as is the case of L$_{\rm z}$, some scattered points
appear in the dispersion plot ($\sigma$$_{\rm V_{LOS}}$,$\sigma$$_{\rm V_{Tan}}$ versus
$\sigma$$_{\rm L_z}$ diagrams). If we discarded these few points, a clearer linear relation
between the properties involved would arise. Since $\sigma$$_{\rm V_{LOS}}$  and
$\sigma$$_{\rm V_{Tan}}$ are well correlated, the scattered points in the
$\sigma$$_{\rm V_{LOS}}$,$\sigma$$_{\rm V_{Tan}}$ versus $\sigma$$_{\rm L_z}$ planes
should be due to  $\sigma$$_{\rm L_z}$. I have checked the fits of L$_{\rm z}$ as a
function of $\phi_1$ for these scatted points (NGC~1851, NGC~1904) and found a
well-fitted function with points slightly departing from the mean fitted
relationship. This means that $\sigma$$_{\rm L_z}$ cannot be smaller as required
for a better linear behavior with $\sigma$$_{\rm V_{LOS}}$ and $\sigma$$_{\rm V_{Tan}}$,
respectively. 

\begin{figure*}
\includegraphics[width=\textwidth]{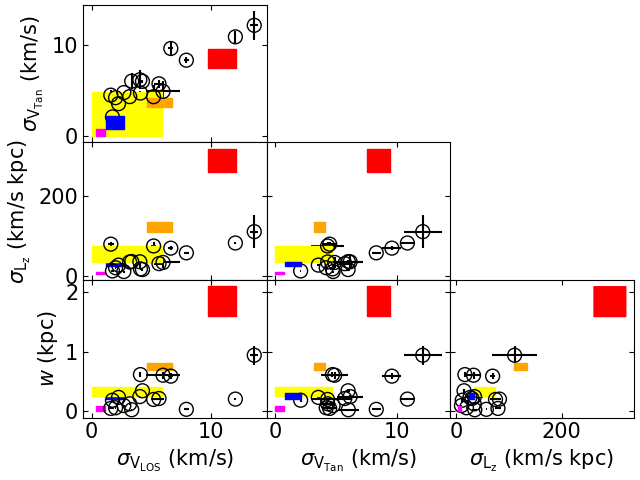}
\caption{Same as Figure~\ref{fig4} with the predictions from \citet{malhanetal2021} 
and \citet{malhanetal2021}, superimposed. Different colors refers to globular
clusters formed in situ (magenta), in 10$^8$$\msun$ and 10$^9$$\msun$ cored profiles
(blue and yellow, respectively), and in 10$^8$$\msun$ and 10$^9$$\msun$ cuspy profiles
(orange and red, respectively).}
\label{fig5}
\end{figure*}

The simulations carried out by \citet{malhanetal2021} and \citet{malhanetal2022}
predicted that globular clusters formed in situ, or in dark matter sub halos 
with cored and cuspy profiles that were accreted into the Milky Way, develop tidal tails 
with increasing width, and dispersion of the z-component of the angular momentum and 
of the line-of-sight and tangential velocities, respectively. I superimposed
the range of values of these properties obtained by \citet{malhanetal2021} and 
\citet{malhanetal2022} on the observation-based results illustrated in Figure~\ref{fig4}.
Figure~\ref{fig5} shows the resulting comparison, where probability distribution
functions of globular clusters formed in situ, in 10$^8$$\msun$ and 10$^9$$\msun$
dark matter halos with cored and cuspy profiles are painted
magenta, blue, yellow, orange and red, respectively.

The simulated properties show steady increasing relationships for all the
examined 2D planes. The values of the four properties for the 10$^9$$\msun$
cuspy profiles are notably the largest ones, while some superposition exits
in  $\sigma$$_{\rm V_{LOS}}$ and $\sigma$$_{\rm V_{Tan}}$ between the 10$^9$$\msun$ 
cored profile and the in situ, 10$^8$$\msun$ cored and 10$^8$$\msun$ cuspy profiles,
respectively. This implies that $\sigma$$_{\rm V_{LOS}}$ and $\sigma$$_{\rm V_{Tan}}$ 
can not unambiguously point to a unique globular cluster's formation scenario, as
it seems to be the case of the simulated $w$ and $\sigma$$_{\rm L_z}$ values. 
Nevertheless, because of the overall monotonic behavior of any paired properties 
seen in Figure~\ref{fig4} and in Figure~\ref{fig5}, the present observation-based 
result bring support to the predicted different origins of globular clusters.

A closer look at Figure~\ref{fig5} reveals that, for each panel, the slopes in 
the observation-based and in the simulated relations are in reasonable good 
agreement for the $\sigma$$_{\rm V_{Tan}}$ -
$\sigma$$_{\rm V_{LOS}}$ and the $\sigma$$_{\rm L_z}$ -  $w$ planes, while
they depart significantly when $\sigma$$_{\rm L_z}$ 
and $w$ are plotted against $\sigma$$_{\rm V_{LOS}}$ or $\sigma$$_{\rm L_z}$.
The reason of such a difference could be due to the predicted much larger values
of  $w$ and $\sigma$$_{\rm L_z}$ for the 10$^9$$\msun$ cuspy model; the
10$^8$$\msun$ cuspy profile also showing some slightly larger $w$ and 
$\sigma$$_{\rm L_z}$ values though. From these behaviors, it would seem to be
recommendable to employ the $\sigma$$_{\rm V_{LOS}}$ versus $\sigma$$_{\rm V_{Tan}}$
plane to infer the origins of globular clusters. I would like to note that
disentangling in the simulations what is causing the split from the observation-based
trends could become a complex analysis of assumptions, different employed parameters,
computational details that are beyond of this work. Indeed, models of globular
clusters' tidal tails based on different physical considerations computed by 
\citet{grondinetal2024} produce different  $w$,  $\sigma$$_{\rm L_z}$, 
$\sigma$$_{\rm V_{LOS}}$, and $\sigma$$_{\rm V_{Tan}}$ relations \citep{piatti2025b}.

\begin{figure}
\includegraphics[width=\columnwidth]{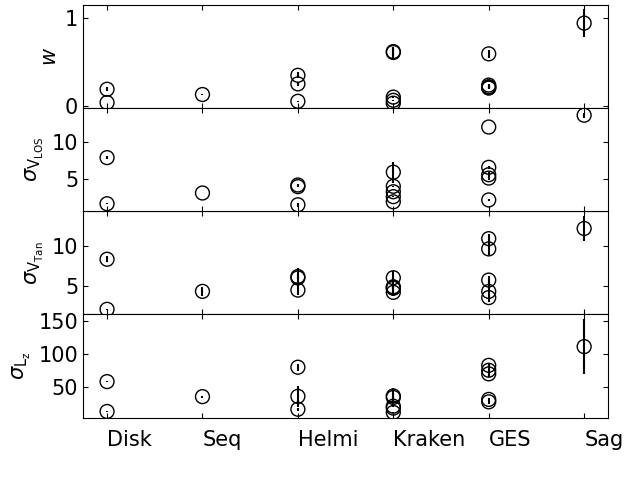}
\caption{$w$ (kpc), V$_{\rm LOS}$ (km/s), V$_{\rm Tan}$ (km/s), and L$_{\rm z}$ (km/s kpc) as 
a function of different accretion groups of the Milky Way, according to \citet{callinghametal2022}, 
namely: Milky Way's disk, Sequoia (Seq), Helmi, Kraken, Gaia-Enceladus-Sausage (GES), and 
Sagittarius (Sag), respectively.}
\label{fig6}
\end{figure}

Finally, I linked the properties of globular cluster's tidal tails of Table~\ref{tab2a} 
with the accretion group of the Milky Way associated to that globular cluster proposed
by \citet{callinghametal2022}. From the defined accretion groups, the presently 
studied globular clusters belong to the Milky Way's disk, Sequoia (Seq), Helmi, Kraken, 
Gaia-Enceladus-Sausage (GES), or Sagittarius (Sag), respectively. Figure~\ref{fig6}
shows the ranges of the parameters' values for these accretion groups. As can be seen, 
tidal tails of globular clusters formed in situ or in an accreted galaxy can be
kinematically cold or hot ($\sigma$ $<$ or $>$ $\sim$ 2 km/s). Bearing in mind the
different formation scenarios of globular clusters proposed by \citet{malhanetal2021} 
and \citet{malhanetal2021}, this implies that the population of globular clusters of a 
galaxy can form within a variety of dark matter halos, in terms of mass and profiles. 
This would seem also valid for globular clusters formed in situ. Besides this overall 
characteristic, Figure~\ref{fig6} also tells us that NGC~6715, the nuclear globular 
cluster of Sagittarius, has the widest and the kinematically hottest tidal tails. 
Although the limited number of globular clusters analyzed, Figure~\ref{fig6} would
still seem to suggest that subtle differences in the ranges of $w$, 
$\sigma$$_{\rm V_{LOS}}$, and $\sigma$$_{\rm V_{Tan}}$, and  $\sigma$$_{\rm L_z}$
exist among different accretion groups of the Milky Way.

\section{Conclusions}

In this study, I explored from an observation-based point of view the scope
of the predictions by \citet{malhanetal2021} and \citet{malhanetal2022} with
respect to the relation claimed between the origin of globular clusters and the
range of values of some properties of their tidal tails, namely: their widths, 
and their dispersion in the z-component of the angular momentum, and in the line-of-sight 
and tangential velocities. I took advantage of the APOGEE DR17 data base in conjunction
with the APOGEE value-added catalog of globular clusters to compile the largest
data set of the four properties mentioned above of globular clusters' tidal tails.
They were computed following the recipes outlines in \citet{malhanetal2021}.
Then, from an in-depth analysis of these properties for 17 globular clusters, the 
following conclusions are drawn:\\

- $\sigma$$_{\rm V_{LOS}}$ and $\sigma$$_{\rm V_{Tan}}$ resulted to be
tightly correlated; correlations involved $\sigma$$_{\rm L_z}$ show a mild
linear relationship, while those with $w$ do not seem to show a robust trend.
However, the latter could be blurred by a number of scattered points. The nearly
identity relationship between  $\sigma$$_{\rm V_{LOS}}$ and $\sigma$$_{\rm V_{Tan}}$ 
reveals that, along the tidal tails, there is not any preference direction of
kinematic agitation. By comparing the resulting trends with those coming from the
probability distribution functions obtained from the simulations carried by
\citet{malhanetal2021} and \citet{malhanetal2022}, I found an overall agreement,
which brings support to their predictions that globular clusters formed in situ or
in dark matter halos with cored or cuspy profiles that were accreted into the Milky
Way span different ranges of values of the four mentioned properties.\\

- Because larger values of $w$ and $\sigma$$_{\rm L_z}$ for dark matter halos with 
cuspy profiles are predicted by the simulations in comparison with the 
observation-based ones obtained here, the $\sigma$$_{\rm V_{LOS}}$ versus 
$\sigma$$_{\rm V_{Tan}}$ plane would seem to be more suitable to infer the
origin of globular clusters. Nevertheless, these latter diagram shows
some superposition between the probability distribution function of the 10$^9$$\msun$
cored dark matter profile with those of in situ, 10$^8$$\msun$ cored and 10$^8$$\msun$ 
cuspy profiles, respectively.\\

- When the values of the four analyzed properties are linked to the accreation groups
of the Milky Way, to which globular clusters are associated, I found kinematically cold 
and hot globular tidal tails belonging to globular clusters distributed in almost all
the considered groups, namely: Milky Way's disk, Sequoia, Helmi, Kraken, 
Gaia-Enceladus-Sausage, and Sagittarius respectively. From the
perspective of the formation scenario of globular clusters, and in the context of 
 \citet{malhanetal2021}'s predictions, this outcome is an observational evidence that
globular clusters can form  within a variety -- in terms of mass and profiles -- of
dark matter halos.

\begin{acknowledgements}

I thank K. Malhan for his kind exchange of ideas.

I thank the referee for the thorough reading of the manuscript and
timely suggestions to improve it.

Data for reproducing the figures and analysis in this work are available upon request
to the first author.
\end{acknowledgements}

%\bibliographystyle{aa}
%\bibliography{paper} % if your bibtex file is called paper.bib

%\input{paper.bbl}

\begin{appendix}
\onecolumn

\section{Selected globular clusters' stars}

\begin{figure*}[h!]
\includegraphics[width=0.3\columnwidth]{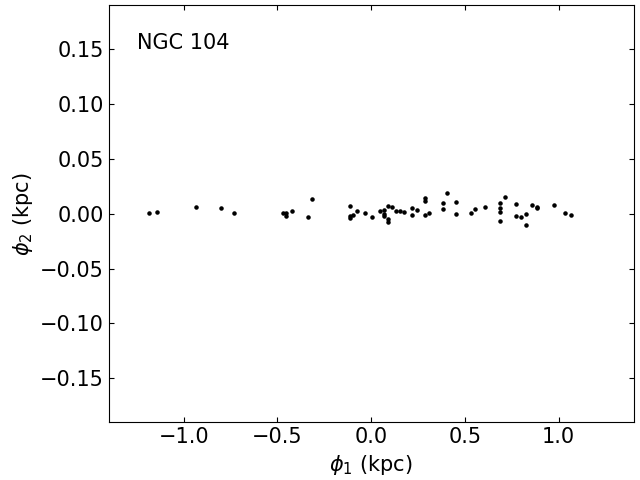}
\includegraphics[width=0.3\columnwidth]{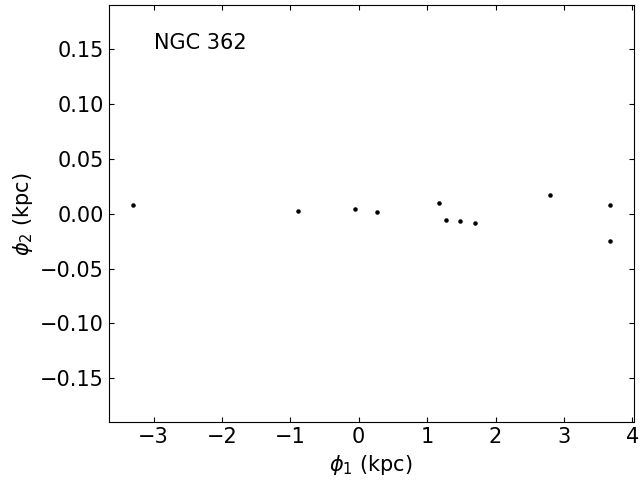}
\includegraphics[width=0.3\columnwidth]{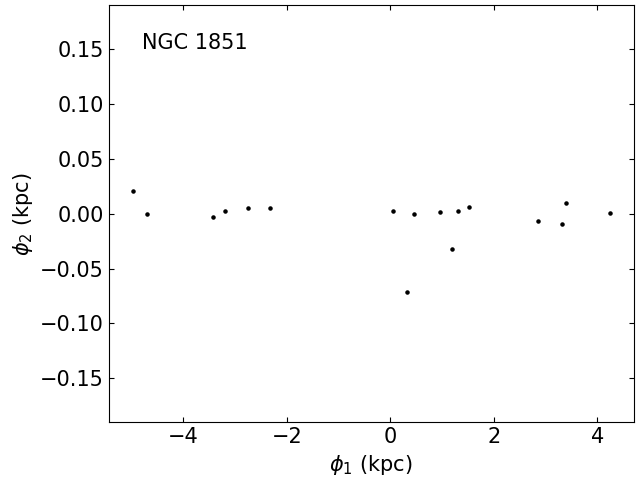}
\includegraphics[width=0.3\columnwidth]{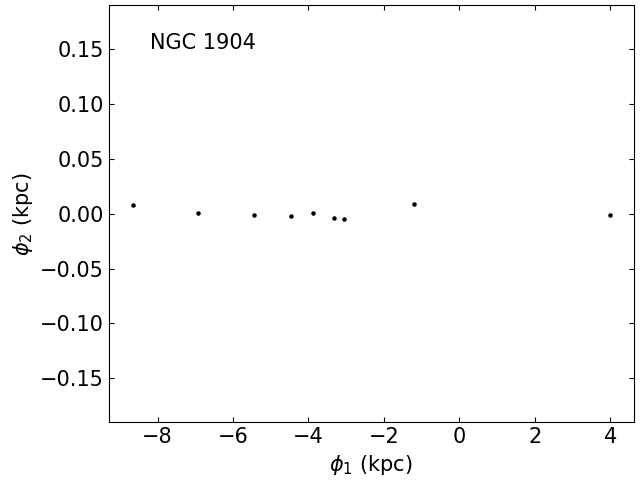}
\includegraphics[width=0.3\columnwidth]{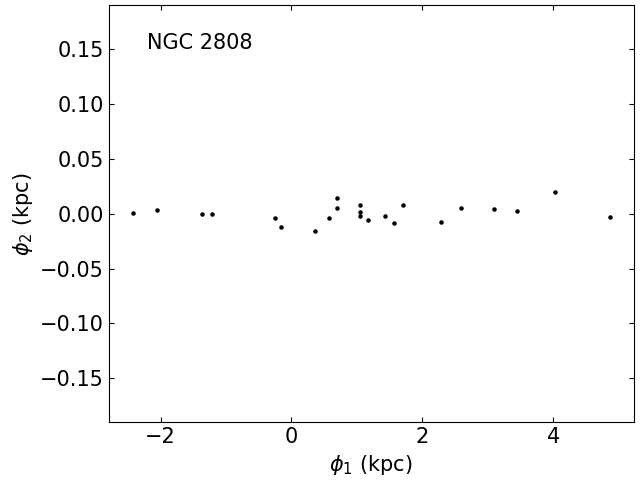}
\includegraphics[width=0.3\columnwidth]{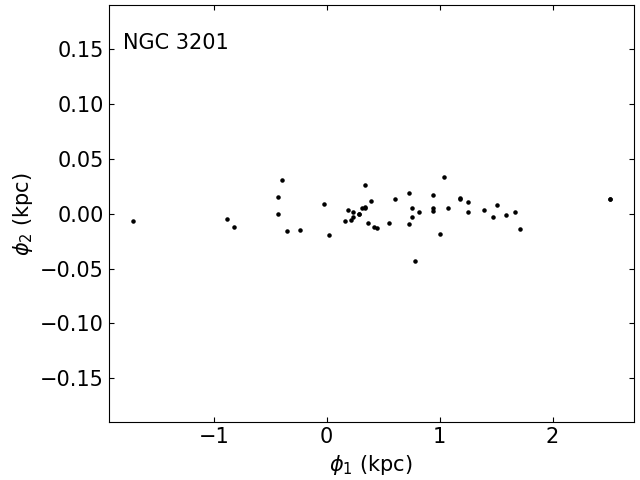}
\includegraphics[width=0.3\columnwidth]{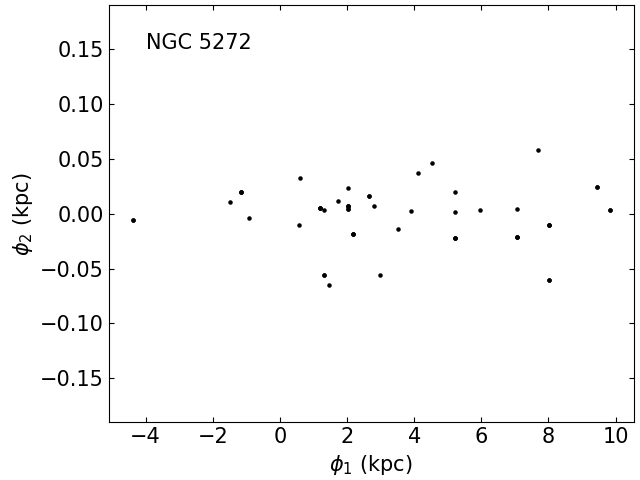}
\includegraphics[width=0.3\columnwidth]{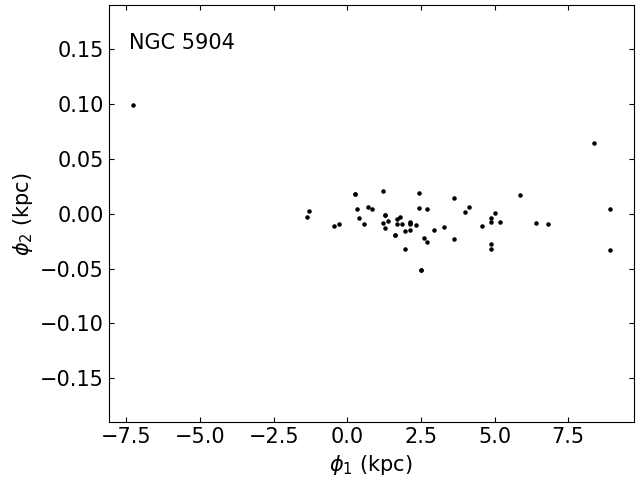}
\includegraphics[width=0.3\columnwidth]{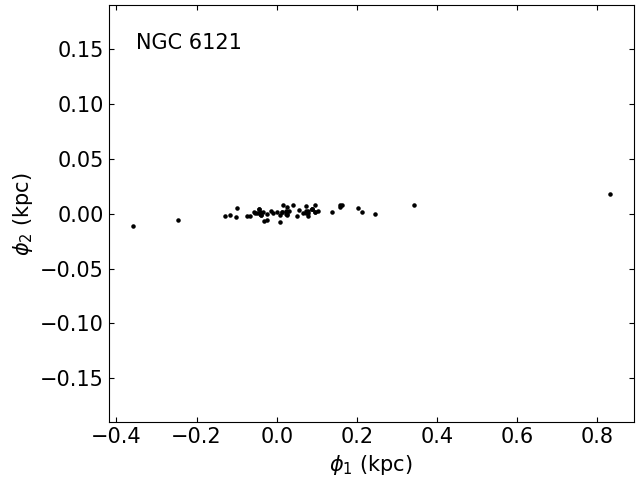}
\includegraphics[width=0.3\columnwidth]{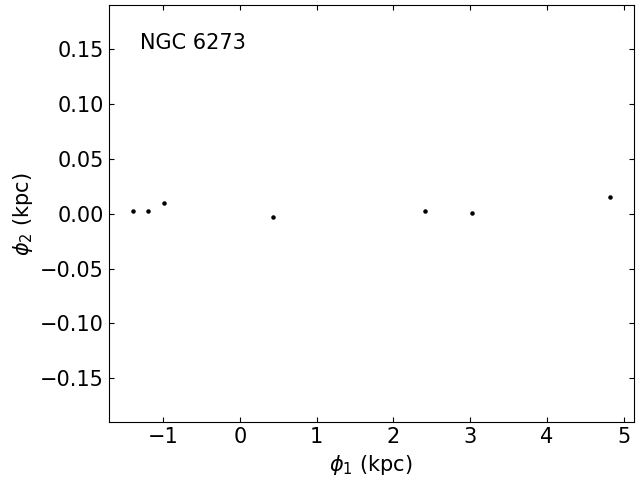}
\includegraphics[width=0.3\columnwidth]{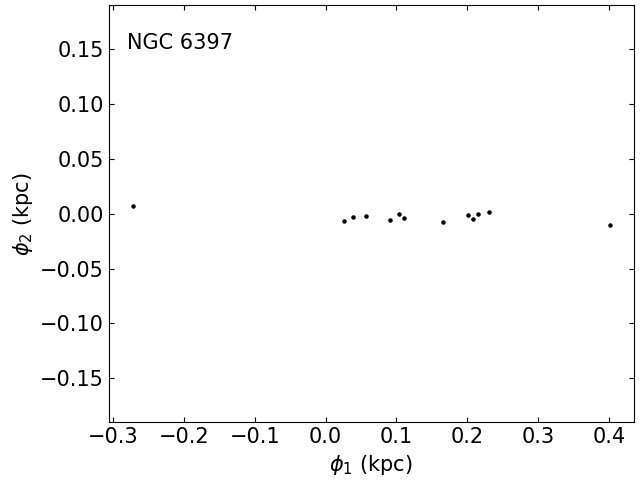}
\includegraphics[width=0.3\columnwidth]{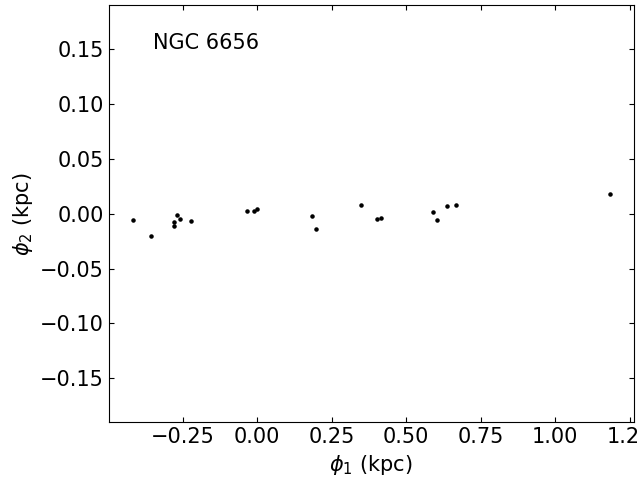}
\includegraphics[width=0.3\columnwidth]{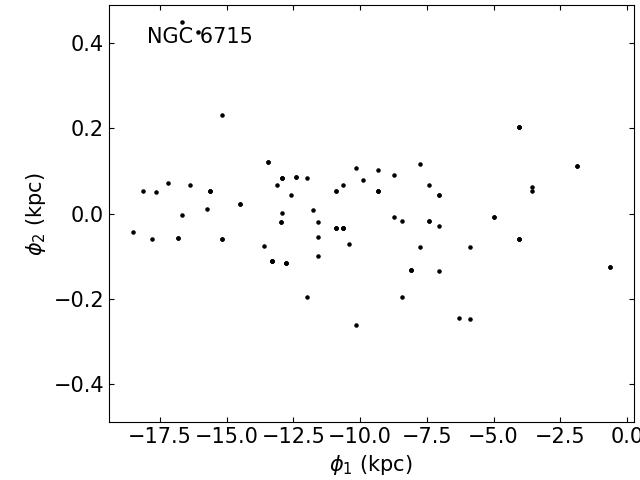}
\includegraphics[width=0.3\columnwidth]{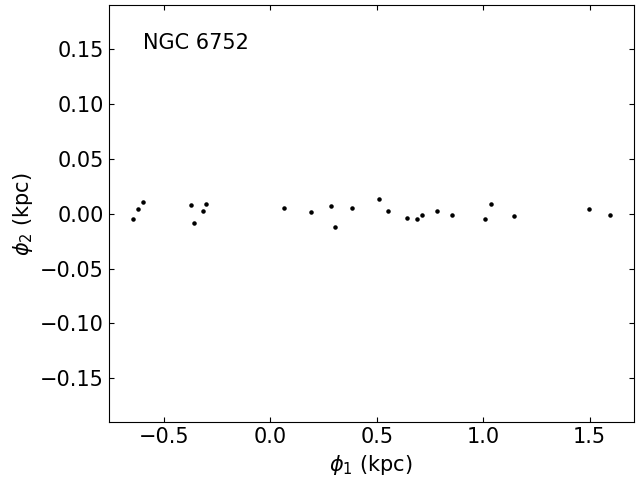}
\includegraphics[width=0.3\columnwidth]{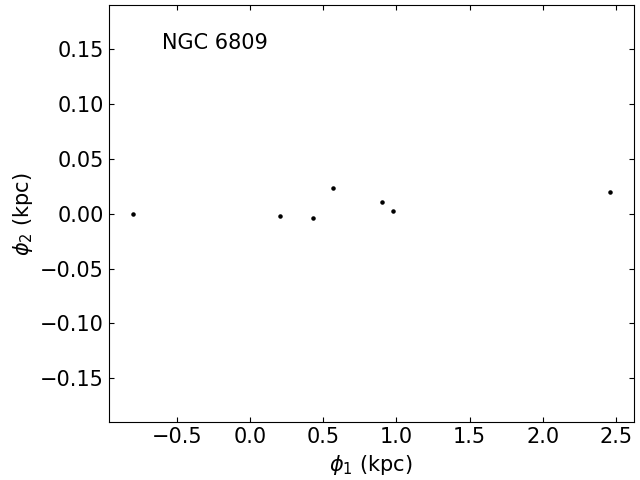}
\includegraphics[width=0.3\columnwidth]{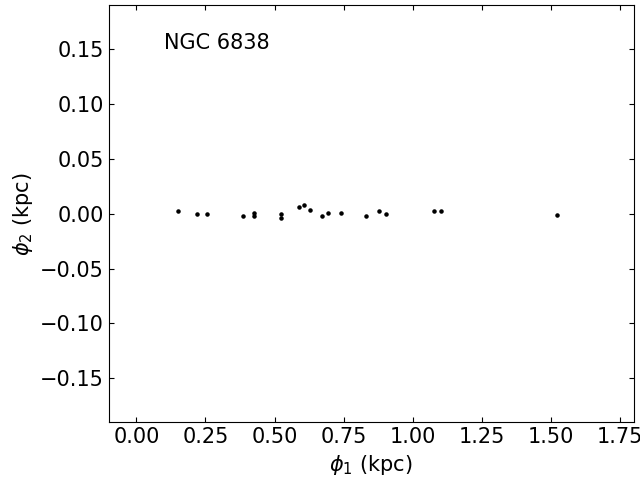}
\caption{Spatial diatribution in the ($\phi_1$,$\phi_2$) plane \citep[see][]{piatti2025b}
of selected tidal tails stars.}
\label{fig2append}
\end{figure*}

\section{Mean properties of Milky Way globular clusters in \citet{schiavonetal2024}'s
APOGEE added-value catalog}

\begin{table}[h!]
\caption{Mean abundances of chemical elements in Milky Way globular clusters.}
\label{tab1a}
\small
\begin{tabular}{lccccccccc}
\hline\hline
Name &  [C/FE] &  [N/FE] & [O/FE] & [MG/FE] & [AL/FE] & [SI/FE] &  [MN/FE] &  [FE/H] & [NI/FE] \\
     &   (dex) &   (dex) & (dex)  & (dex)   &  (dex)  &  (dex)  &   (dex)  &   (dex) &  (dex)  \\
\hline
Djorg~2  & -0.28$\pm$0.11 & 0.62$\pm$0.42 & 0.32$\pm$0.04 & 0.32$\pm$0.04 & 0.21$\pm$0.19 & 0.25$\pm$0.03 &-0.27$\pm$0.02 &-1.08$\pm$0.06 & 0.00$\pm$0.01\\
FSR~1758 & -0.38$\pm$0.18 & 0.76$\pm$0.41 & 0.33$\pm$0.11 & 0.26$\pm$0.09 & 0.15$\pm$0.23 & 0.26$\pm$0.04 &-0.40$\pm$0.23 &-1.43$\pm$0.09 & 0.00$\pm$0.11\\
HP~1     & -0.41$\pm$0.12 & 0.93$\pm$0.30 & 0.27$\pm$0.09 & 0.17$\pm$0.06 & 0.29$\pm$0.29 & 0.23$\pm$0.03 &-0.25$\pm$0.18 &-1.21$\pm$0.15 &-0.02$\pm$0.08\\
Liller~1 &  0.05$\pm$0.09 & 0.29$\pm$0.19 & 0.12$\pm$0.11 & 0.09$\pm$0.13 &-0.05$\pm$0.03 & 0.03$\pm$0.09 & 0.62$\pm$0.01 &-0.16$\pm$0.39 & 0.11$\pm$0.05\\
NGC~ 104 & -0.03$\pm$0.16 & 0.64$\pm$0.40 & 0.31$\pm$0.12 & 0.33$\pm$0.04 & 0.31$\pm$0.13 & 0.24$\pm$0.04 &-0.18$\pm$0.10 &-0.75$\pm$0.06 & 0.07$\pm$0.04\\
NGC~ 288 & -0.22$\pm$0.09 & 0.48$\pm$0.32 & 0.40$\pm$0.10 & 0.27$\pm$0.04 & 0.15$\pm$0.14 & 0.30$\pm$0.03 &-0.36$\pm$0.15 &-1.28$\pm$0.06 & 0.00$\pm$0.04\\
NGC~ 362 & -0.44$\pm$0.18 & 0.55$\pm$0.40 & 0.17$\pm$0.15 & 0.08$\pm$0.07 &-0.00$\pm$0.24 & 0.12$\pm$0.04 &-0.30$\pm$0.14 &-1.11$\pm$0.06 &-0.05$\pm$0.09\\
NGC~1851 & -0.32$\pm$0.30 & 0.55$\pm$0.57 & 0.23$\pm$0.22 & 0.21$\pm$0.14 &-0.04$\pm$0.34 & 0.17$\pm$0.13 &-0.26$\pm$0.24 &-1.14$\pm$0.17 & 0.07$\pm$0.27\\
NGC~1904 & -0.62$\pm$0.19 & 0.67$\pm$0.31 & 0.14$\pm$0.17 & 0.08$\pm$0.11 & 0.13$\pm$0.46 & 0.15$\pm$0.06 &-0.38$\pm$0.19 &-1.52$\pm$0.13 &-0.02$\pm$0.25\\
NGC~2298 & -0.43$\pm$0.22 & 0.69$\pm$0.33 & 0.21$\pm$0.22 & 0.14$\pm$0.13 & 0.03$\pm$0.47 & 0.24$\pm$0.03 &-0.25$\pm$0.18 &-1.84$\pm$0.07 &-0.04$\pm$0.11\\
NGC~2808 & -0.43$\pm$0.11 & 0.59$\pm$0.38 & 0.14$\pm$0.18 & 0.04$\pm$0.17 & 0.21$\pm$0.45 & 0.15$\pm$0.05 &-0.23$\pm$0.13 &-1.07$\pm$0.07 &-0.01$\pm$0.08\\
NGC~3201 & -0.15$\pm$0.27 & 0.24$\pm$0.50 & 0.15$\pm$0.27 & 0.16$\pm$0.11 &-0.17$\pm$0.33 & 0.15$\pm$0.06 &-0.10$\pm$0.34 &-1.39$\pm$0.12 &-0.05$\pm$0.17\\
NGC~4147 & -0.54$\pm$0.07 & 0.49$\pm$0.27 & 0.33$\pm$0.01 & 0.24$\pm$0.03 & 0.05$\pm$0.34 & 0.21$\pm$0.01 &-0.12$\pm$0.32 &-1.62$\pm$0.06 &-0.06$\pm$0.05\\
NGC~4590 & -0.42$\pm$0.38 & 0.23$\pm$0.48 & 0.31$\pm$0.16 & 0.26$\pm$0.10 & 0.14$\pm$0.37 & 0.31$\pm$0.05 & 0.07$\pm$0.18 &-2.22$\pm$0.08 & 0.16$\pm$0.21\\
NGC~5024 & -0.35$\pm$0.24 & 0.65$\pm$0.22 & 0.32$\pm$0.09 & 0.25$\pm$0.09 & 0.12$\pm$0.41 & 0.24$\pm$0.06 &-0.18$\pm$0.17 &-1.90$\pm$0.09 & 0.00$\pm$0.15\\
NGC~5053 & -0.06$\pm$0.36 & 0.35$\pm$0.40 & 0.37$\pm$0.25 & 0.25$\pm$0.12 & 0.27$\pm$0.41 & 0.32$\pm$0.06 & 0.08$\pm$0.25 &-2.20$\pm$0.11 &-0.05$\pm$0.18\\
NGC~5139 & -0.04$\pm$0.32 & 0.61$\pm$0.54 & 0.32$\pm$0.24 & 0.23$\pm$0.20 & 0.26$\pm$0.51 & 0.29$\pm$0.07 &-0.32$\pm$0.25 &-1.60$\pm$0.24 &-0.01$\pm$0.13\\
NGC~5272 & -0.32$\pm$0.25 & 0.29$\pm$0.41 & 0.27$\pm$0.18 & 0.13$\pm$0.08 &-0.05$\pm$0.33 & 0.18$\pm$0.10 &-0.40$\pm$0.18 &-1.42$\pm$0.10 &-0.12$\pm$0.11\\
NGC~5466 & -0.64$\pm$0.31 & 0.46$\pm$0.37 & 0.17$\pm$0.07 & 0.12$\pm$0.10 &-0.27$\pm$0.30 & 0.12$\pm$0.08 &-0.19$\pm$0.11 &-1.81$\pm$0.08 &-0.11$\pm$0.21\\
NGC~5634 & -0.62$\pm$0.04 & 0.56$\pm$0.19 & 0.34$\pm$0.04 & 0.21$\pm$0.06 & 0.00$\pm$0.32 & 0.20$\pm$0.04 &-0.35$\pm$0.06 &-1.72$\pm$0.05 & 0.01$\pm$0.03\\
NGC~5904 & -0.27$\pm$0.20 & 0.61$\pm$0.42 & 0.26$\pm$0.23 & 0.15$\pm$0.08 & 0.07$\pm$0.36 & 0.18$\pm$0.06 &-0.37$\pm$0.22 &-1.20$\pm$0.10 &-0.07$\pm$0.20\\
NGC~6093 & -0.58$\pm$0.01 & 0.38$\pm$0.01 & 0.36$\pm$0.06 & 0.23$\pm$0.03 &-0.14$\pm$0.15 & 0.31$\pm$0.01 &-0.22$\pm$0.18 &-1.60$\pm$0.01 & 0.02$\pm$0.06\\
NGC~6121 & -0.05$\pm$0.11 & 0.62$\pm$0.45 & 0.40$\pm$0.10 & 0.34$\pm$0.05 & 0.41$\pm$0.12 & 0.37$\pm$0.05 &-0.22$\pm$0.12 &-1.07$\pm$0.10 & 0.06$\pm$0.06\\
NGC~6144 & -1.02$\pm$0.19 & 1.98$\pm$0.41 & 0.70$\pm$0.15 & 0.53$\pm$0.12 &-0.28$\pm$0.07 & 0.43$\pm$0.12 &     ---       &-1.82$\pm$0.05 &   ---        \\
NGC~6171 & -0.02$\pm$0.11 & 0.68$\pm$0.43 & 0.44$\pm$0.17 & 0.31$\pm$0.06 & 0.27$\pm$0.13 & 0.33$\pm$0.04 &-0.32$\pm$0.14 &-1.01$\pm$0.10 & 0.04$\pm$0.09\\
NGC~6205 & -0.34$\pm$0.25 & 0.70$\pm$0.38 & 0.15$\pm$0.23 & 0.06$\pm$0.14 & 0.38$\pm$0.44 & 0.20$\pm$0.07 &-0.33$\pm$0.25 &-1.48$\pm$0.09 &-0.02$\pm$0.13\\
NGC~6218 & -0.15$\pm$0.20 & 0.54$\pm$0.40 & 0.31$\pm$0.13 & 0.30$\pm$0.05 & 0.07$\pm$0.17 & 0.26$\pm$0.04 &-0.49$\pm$0.23 &-1.27$\pm$0.06 &-0.02$\pm$0.14\\
NGC~6229 & -0.48$\pm$0.08 & 0.44$\pm$0.32 & 0.19$\pm$0.12 & 0.14$\pm$0.06 &-0.08$\pm$0.25 & 0.17$\pm$0.10 &-0.27$\pm$0.03 &-1.24$\pm$0.06 &-0.10$\pm$0.07\\
NGC~6254 & -0.36$\pm$0.17 & 0.49$\pm$0.47 & 0.25$\pm$0.18 & 0.21$\pm$0.11 & 0.20$\pm$0.46 & 0.24$\pm$0.04 &-0.17$\pm$0.16 &-1.51$\pm$0.07 &-0.01$\pm$0.07\\
NGC~6273 & -0.25$\pm$0.33 & 0.84$\pm$0.37 & 0.18$\pm$0.16 & 0.17$\pm$0.11 & 0.35$\pm$0.41 & 0.22$\pm$0.07 &-0.16$\pm$0.23 &-1.71$\pm$0.13 & 0.04$\pm$0.16\\
NGC~6293 & -0.57$\pm$0.40 & 0.68$\pm$0.39 & 0.21$\pm$0.23 & 0.13$\pm$0.16 & 0.37$\pm$0.36 & 0.30$\pm$0.08 & 0.01$\pm$0.20 &-2.09$\pm$0.07 & 0.06$\pm$0.20\\
NGC~6304 & -0.09$\pm$0.24 & 0.66$\pm$0.42 & 0.25$\pm$0.10 & 0.28$\pm$0.05 & 0.30$\pm$0.11 & 0.20$\pm$0.03 &-0.09$\pm$0.05 &-0.48$\pm$0.05 & 0.08$\pm$0.04\\
NGC~6316 & -0.08$\pm$0.11 & 0.59$\pm$0.37 & 0.28$\pm$0.03 & 0.28$\pm$0.05 & 0.22$\pm$0.15 & 0.25$\pm$0.05 &-0.05$\pm$0.06 &-0.76$\pm$0.04 & 0.07$\pm$0.03\\
NGC~6341 & -0.09$\pm$0.42 & 1.04$\pm$0.47 & 0.33$\pm$0.29 & 0.15$\pm$0.20 & 0.35$\pm$0.33 & 0.34$\pm$0.07 & 0.24$\pm$0.19 &-2.24$\pm$0.07 & 0.09$\pm$0.19\\
NGC~6380 & -0.15$\pm$0.17 & 0.78$\pm$0.42 & 0.26$\pm$0.09 & 0.29$\pm$0.05 & 0.43$\pm$0.26 & 0.22$\pm$0.07 &-0.15$\pm$0.09 &-0.76$\pm$0.18 & 0.07$\pm$0.03\\
NGC~6388 & -0.12$\pm$0.15 & 0.63$\pm$0.39 & 0.08$\pm$0.07 & 0.06$\pm$0.07 & 0.18$\pm$0.26 & 0.05$\pm$0.09 &-0.00$\pm$0.07 &-0.48$\pm$0.19 & 0.00$\pm$0.06\\
NGC~6397 &  0.07$\pm$0.32 & 0.58$\pm$0.46 & 0.30$\pm$0.24 & 0.28$\pm$0.07 & 0.26$\pm$0.36 & 0.28$\pm$0.05 & 0.05$\pm$0.23 &-2.02$\pm$0.06 & 0.05$\pm$0.18\\
NGC~6401 & -0.03$\pm$0.13 & 0.38$\pm$0.74 & 0.25$\pm$0.20 & 0.16$\pm$0.06 &-0.12$\pm$0.37 & 0.23$\pm$0.16 &-0.13$\pm$0.67 &-1.07$\pm$0.15 & 0.47$\pm$0.53\\
NGC~6441 & -0.00$\pm$0.12 & 0.49$\pm$0.32 & 0.21$\pm$0.07 & 0.21$\pm$0.06 & 0.23$\pm$0.27 & 0.16$\pm$0.12 &-0.06$\pm$0.07 &-0.47$\pm$0.18 & 0.03$\pm$0.07\\
NGC~6517 & -0.42$\pm$0.08 & 0.20$\pm$0.07 & 0.39$\pm$0.07 & 0.24$\pm$0.04 &-0.31$\pm$0.05 & 0.25$\pm$0.04 &-0.16$\pm$0.07 &-1.58$\pm$0.02 &-0.24$\pm$0.06\\
NGC~6522 & -0.30$\pm$0.19 & 0.79$\pm$0.37 & 0.33$\pm$0.22 & 0.15$\pm$0.09 & 0.23$\pm$0.25 & 0.22$\pm$0.04 &-0.22$\pm$0.08 &-1.22$\pm$0.11 &-0.01$\pm$0.18\\
NGC~6528 &  0.05$\pm$0.03 & 0.49$\pm$0.24 & 0.17$\pm$0.02 & 0.17$\pm$0.04 & 0.03$\pm$0.02 & 0.05$\pm$0.02 &-0.01$\pm$0.02 &-0.15$\pm$0.01 & 0.06$\pm$0.03\\
NGC~6539 &  0.02$\pm$0.03 & 0.01$\pm$0.03 & 0.42$\pm$0.03 & 0.37$\pm$0.02 & 0.34$\pm$0.04 & 0.21$\pm$0.03 &-0.27$\pm$0.04 &-0.74$\pm$0.01 & 0.04$\pm$0.03\\
NGC~6540 & -0.32$\pm$0.18 & 0.80$\pm$0.40 & 0.22$\pm$0.13 & 0.22$\pm$0.08 & 0.18$\pm$0.10 & 0.22$\pm$0.04 &-0.37$\pm$0.08 &-1.02$\pm$0.03 & 0.00$\pm$0.03\\
NGC~6544 & -0.20$\pm$0.21 & 0.41$\pm$0.38 & 0.27$\pm$0.12 & 0.20$\pm$0.06 & 0.04$\pm$0.23 & 0.23$\pm$0.04 &-0.32$\pm$0.23 &-1.51$\pm$0.09 &-0.04$\pm$0.08\\
NGC~6553 & -0.12$\pm$0.17 & 0.77$\pm$0.30 & 0.18$\pm$0.06 & 0.21$\pm$0.03 & 0.18$\pm$0.06 & 0.09$\pm$0.03 & 0.03$\pm$0.04 &-0.18$\pm$0.02 & 0.07$\pm$0.02\\
NGC~6558 & -0.31$\pm$0.05 & 0.84$\pm$0.33 & 0.22$\pm$0.15 & 0.21$\pm$0.12 & 0.26$\pm$0.25 & 0.12$\pm$0.20 &-0.34$\pm$0.15 &-1.00$\pm$0.21 & 0.05$\pm$0.19\\
NGC~6569 & -0.18$\pm$0.15 & 0.58$\pm$0.38 & 0.29$\pm$0.08 & 0.31$\pm$0.04 & 0.28$\pm$0.16 & 0.29$\pm$0.07 &-0.21$\pm$0.04 &-0.98$\pm$0.03 & 0.03$\pm$0.02\\
NGC~6642 & -0.18$\pm$0.06 & 0.54$\pm$0.33 & 0.33$\pm$0.05 & 0.31$\pm$0.03 & 0.21$\pm$0.10 & 0.30$\pm$0.05 &-0.28$\pm$0.07 &-1.09$\pm$0.03 & 0.04$\pm$0.04\\
NGC~6656 & -0.07$\pm$0.33 & 0.63$\pm$0.48 & 0.31$\pm$0.19 & 0.25$\pm$0.09 & 0.30$\pm$0.40 & 0.26$\pm$0.06 &-0.34$\pm$0.25 &-1.70$\pm$0.10 &-0.07$\pm$0.16\\
NGC~6715 & -0.33$\pm$0.16 & 0.09$\pm$0.24 &-0.00$\pm$0.09 &-0.04$\pm$0.09 &-0.36$\pm$0.29 &-0.07$\pm$0.10 &-0.16$\pm$0.11 &-0.63$\pm$0.36 &-0.12$\pm$0.08\\
NGC~6717 & -0.23$\pm$0.04 & 0.36$\pm$0.41 & 0.34$\pm$0.08 & 0.32$\pm$0.01 &-0.05$\pm$0.12 & 0.23$\pm$0.01 &-0.55$\pm$0.23 &-1.13$\pm$0.04 & 0.00$\pm$0.02\\
NGC~6723 & -0.29$\pm$0.09 & 0.69$\pm$0.37 & 0.28$\pm$0.05 & 0.26$\pm$0.05 & 0.16$\pm$0.13 & 0.25$\pm$0.03 &-0.24$\pm$0.03 &-1.03$\pm$0.05 & 0.02$\pm$0.02\\
NGC~6752 & -0.07$\pm$0.22 & 0.56$\pm$0.46 & 0.18$\pm$0.27 & 0.19$\pm$0.11 & 0.36$\pm$0.44 & 0.23$\pm$0.05 &-0.38$\pm$0.23 &-1.48$\pm$0.15 & 0.01$\pm$0.09\\
NGC~6760 & -0.01$\pm$0.10 & 0.76$\pm$0.29 & 0.30$\pm$0.02 & 0.29$\pm$0.02 & 0.20$\pm$0.09 & 0.20$\pm$0.02 &-0.09$\pm$0.02 &-0.74$\pm$0.04 & 0.12$\pm$0.02\\
NGC~6809 & -0.48$\pm$0.30 & 0.73$\pm$0.37 & 0.23$\pm$0.18 & 0.26$\pm$0.08 & 0.20$\pm$0.40 & 0.23$\pm$0.05 &-0.26$\pm$0.17 &-1.76$\pm$0.07 & 0.04$\pm$0.08\\
NGC~6838 &  0.01$\pm$0.11 & 0.37$\pm$0.38 & 0.35$\pm$0.13 & 0.32$\pm$0.04 & 0.21$\pm$0.07 & 0.24$\pm$0.06 &-0.22$\pm$0.09 &-0.74$\pm$0.09 & 0.07$\pm$0.03\\
NGC~7078 & -0.22$\pm$0.49 & 0.74$\pm$0.56 & 0.31$\pm$0.34 & 0.16$\pm$0.25 & 0.30$\pm$0.34 & 0.35$\pm$0.08 & 0.31$\pm$0.31 &-2.28$\pm$0.09 & 0.12$\pm$0.25\\
NGC~7089 & -0.45$\pm$0.18 & 0.56$\pm$0.26 & 0.22$\pm$0.18 & 0.10$\pm$0.08 & 0.14$\pm$0.40 & 0.18$\pm$0.04 &-0.31$\pm$0.09 &-1.47$\pm$0.05 &-0.06$\pm$0.07\\
\hline
\end{tabular}
\end{table}

\setcounter{table}{0}
\begin{table}
\caption{continued.}
\label{tab1a}
\small
\begin{tabular}{lccccccccc}
\hline\hline
Name &  [C/FE] &  [N/FE] & [O/FE] & [MG/FE] & [AL/FE] & [SI/FE] &  [MN/FE] &  [FE/H] & [NI/FE] \\
     &   (dex) &   (dex) & (dex)  & (dex)   &  (dex)  &  (dex)  &   (dex)  &   (dex) &  (dex)  \\
\hline
Pal~1    &  0.27$\pm$0.12 & 0.41$\pm$0.08 & 0.52$\pm$0.04 & 0.01$\pm$0.07 & 0.03$\pm$0.09 & 0.12$\pm$0.02 &-0.04$\pm$0.03 &-0.44$\pm$0.01 &-0.02$\pm$0.04\\
Pal~5    & -0.35$\pm$0.02 & 0.38$\pm$0.18 & 0.28$\pm$0.09 & 0.12$\pm$0.03 &-0.14$\pm$0.07 & 0.14$\pm$0.02 &-0.34$\pm$0.02 &-1.24$\pm$0.05 &-0.04$\pm$0.01\\
Pal~6    & -0.17$\pm$0.11 & 0.47$\pm$0.31 & 0.31$\pm$0.05 & 0.31$\pm$0.03 & 0.15$\pm$0.06 & 0.28$\pm$0.08 &-0.15$\pm$0.03 &-0.92$\pm$0.08 & 0.06$\pm$0.04\\
Pal~10   &  0.03$\pm$0.01 & 0.21$\pm$0.01 & 0.07$\pm$0.01 & 0.04$\pm$0.01 &    ---        &-0.03$\pm$0.01 &   ---         & 0.01$\pm$0.02 & 0.02$\pm$0.01\\
Rup~106  & -0.59$\pm$0.10 &-0.00$\pm$0.16 & 0.10$\pm$0.06 &-0.14$\pm$0.03 &-0.62$\pm$0.04 &-0.05$\pm$0.02 &-0.61$\pm$0.13 &-1.30$\pm$0.03 &-0.11$\pm$0.02\\
Ter~2    & -0.08$\pm$0.05 & 0.56$\pm$0.33 & 0.31$\pm$0.01 & 0.35$\pm$0.03 & 0.12$\pm$0.11 & 0.24$\pm$0.02 &-0.08$\pm$0.01 &-0.85$\pm$0.04 & 0.04$\pm$0.01\\
Ter~4    & -0.42$\pm$0.01 & 0.59$\pm$0.28 & 0.20$\pm$0.12 & 0.14$\pm$0.12 & 0.18$\pm$0.43 & 0.21$\pm$0.03 &-0.31$\pm$0.03 &-1.38$\pm$0.06 &-0.03$\pm$0.07\\
Ter~9    & -0.38$\pm$0.24 & 0.65$\pm$0.34 & 0.26$\pm$0.25 & 0.19$\pm$0.10 & 0.12$\pm$0.43 & 0.18$\pm$0.05 &-0.25$\pm$0.19 &-1.37$\pm$0.05 & 0.03$\pm$0.10\\
Ter~10   & -0.69$\pm$0.04 & 0.94$\pm$0.04 & 0.25$\pm$0.03 & 0.23$\pm$0.03 & 0.49$\pm$0.03 & 0.21$\pm$0.03 &-0.27$\pm$0.05 &-1.62$\pm$0.01 & 0.06$\pm$0.04\\
Ter~12   &  0.13$\pm$0.01 & 0.46$\pm$0.01 & 0.33$\pm$0.01 & 0.33$\pm$0.01 & 0.03$\pm$0.02 & 0.20$\pm$0.01 &  ---          &-0.56$\pm$0.01 & 0.14$\pm$0.01\\
Ton~2    & -0.07$\pm$0.10 & 0.73$\pm$0.34 & 0.26$\pm$0.09 & 0.30$\pm$0.07 & 0.39$\pm$0.08 & 0.21$\pm$0.09 &-0.19$\pm$0.08 &-0.74$\pm$0.28 & 0.06$\pm$0.03\\
UKS~1    & -0.28$\pm$0.21 & 0.67$\pm$0.32 & 0.19$\pm$0.14 & 0.17$\pm$0.21 & 0.13$\pm$0.13 & 0.18$\pm$0.17 &-0.04$\pm$0.05 &-0.99$\pm$0.17 & 0.03$\pm$0.06\\
\hline
\end{tabular}
\end{table}

\begin{table}[h!]
\caption{$\sigma$V$_{\rm LOS}$, $\sigma$V$_{\rm Tan}$, and $\sigma$L$_{\rm z}$ values
of Milky Way globular clusters.}
\label{tab2a}
%\small
\begin{tabular}{lcccccc}
\hline\hline
Name & N & length & $w$ & $\sigma$V$_{\rm LOS}$ & $\sigma$V$_{\rm Tan}$ & $\sigma$L$_{\rm z}$ \\
     &   & (kpc) &  (pc) &(km/s)               &      (km/s)           &  (km/s kpc) \\
\hline
NGC~104  & 60 & 9.3 & 5.6$\pm$0.1 &  7.93$\pm$0.21 &  8.30$\pm$0.35 & 57.63$\pm$0.79  \\
NGC~362  & 11& 7.4 & 39.0$\pm$3.0 &  2.26$\pm$0.12 &  3.55$\pm$0.14 & 26.75$\pm$2.59  \\
NGC~1851 & 20&14.1 & 33.3$\pm$2.1 &  5.19$\pm$0.20 &  4.31$\pm$1.32 & 74.93$\pm$10.34 \\
NGC~1904 & 9& 6.1 & 8.1$\pm$1.0 &  1.61$\pm$0.23 &  4.48$\pm$0.57 & 79.56$\pm$5.12 \\
NGC~2808 & 23& 6.4 & 98.4$\pm$8.2 &  6.63$\pm$0.20 &  9.58$\pm$0.78 & 69.45$\pm$5.30\\                      
NGC~3201 & 51& 4.2 & 21.0$\pm$1.0 &  3.19$\pm$0.06 &  4.32$\pm$0.52 & 34.69$\pm$1.69 \\
NGC~5139 & 160& 6.6 & 34.3$\pm$2.1 & 12.03$\pm$0.06 & 10.85$\pm$0.63 & 82.44$\pm$2.07 \\
NGC~5272 & 58&14.2 & 41.3$\pm$3.5 &  4.04$\pm$0.11 &  6.16$\pm$1.05 & 35.14$\pm$16.45 \\
NGC~5904 & 50& 6.2 & 57.6$\pm$5.7 &  4.26$\pm$0.08 &  5.99$\pm$0.15 & 15.49$\pm$2.49 \\
NGC~6121 & 61& 1.2 & 4.5$\pm$0.2 &  3.36$\pm$0.03 &  6.01$\pm$0.87 & 35.76$\pm$4.62 \\
NGC~6273 & 7& 5.3 & 101.7$\pm$9.6 &  5.98$\pm$1.39 &  4.89$\pm$1.09 & 33.44$\pm$14.16 \\
NGC~6397 & 13& 0.7 & 16.1$\pm$0.9 &  2.70$\pm$0.07 &  4.76$\pm$0.08 & 10.58$\pm$0.70 \\
NGC~6656 & 20& 1.6 & 35.9$\pm$2.4 &  5.64$\pm$0.38 &  5.72$\pm$0.47 & 30.45$\pm$2.93 \\
NGC~6715 & 113&17.9 & 156.9$\pm$26.0 & 13.61$\pm$0.34 & 12.11$\pm$1.56 &110.99$\pm$41.98 \\
NGC~6752 & 24& 2.2 & 102.8$\pm$8.0 &  4.09$\pm$0.05 &  4.71$\pm$0.21 & 17.17$\pm$0.45  \\    
NGC~6809 & 7& 3.2 & 10.0$\pm$0.6 &  2.01$\pm$0.02 &  4.20$\pm$0.14 & 20.09$\pm$1.15 \\
NGC~6838 & 20&1.4 & 31.1$\pm$4.0 &  1.75$\pm$0.10 &  2.10$\pm$0.05 & 12.07$\pm$0.74 \\
\hline
\end{tabular}
\end{table}

\begin{table}[h!]
\caption{Source\_ID from APOGEE DR17 of selected stars of globular clusters.}
\label{tab3a}
%\small
\begin{tabular}{l}
\hline
NGC~104: \\4689839858983442432, 4689833124475021312, 4689831509567399936
 4689807766988459008, 4689806701836580864, \\

4689651086596023296,
 4689646997787297792, 4689646241872961536, 4689646070074297344,
 4689645928342350848, \\

4689644249005617152, 4689644111567130624,
 4689644042847672320, 4689643905408839680, 4689643218214133760,\\

 4689643149494738944, 4689642702818393088, 4689642286190119936,
 4689642251832561664, 4689641946904328192, \\

4689641392838717440,
 4689641190988963840, 4689640984830517248, 4689640946182242304,
 4689640400715718656, \\

4689639988398164992, 4689639885318980608,
 4689639782230993920, 4689639777957340160, 4689639674858442752,\\

 4689639606138968064, 4689639503059722240, 4689639434340312064,
 4689639159462279168, 4689638785808484352,\\

 4689638678425891840,
 4689638442211458048, 4689638403539699712, 4689638334822345728,
 4689638236045838336, \\

4689637995535079424, 4689637991231178752,
 4689637961174887424, 4689637617577709568, 4689637239620660224,\\

 4689633941083703296, 4689633082090281984, 4689632738492989440,
 4689632669773531136, 4689628129980152832, \\

4689627958187766784,
 4689627958187765760, 4689625415574266880, 4689625316788603904,
 4689625144980305920, \\

4689624698313312256, 4689614837066803200,
 4689614802707077120, 4689614733987610624, 4689613458369686528\\
\hline
NGC~362:\\ 4690890854666110976, 4690887590491024384, 4690887384332587008,
 4690887006375453696, 4690886834576765952, \\4690886628410998784,
 4690839899175858176, 4690839727376427008, 4690839723077788672,
 4690839521220458496, \\4690839280692953088\\
\hline
NGC~1851:\\ 4819292612825765888, 4819198570223966208, 4819198432785021952,
 4819197986108430336, 4819197814310430720, \\4819197814309866496,
 4819197779950716928, 4819197706932441088, 4819197676871329792,
 4819197573791676416,\\ 4819197573791662080, 4819197470709227520,
 4819197298914465792, 4819197088457132032, 4819195546567853056,\\
 4819185719681969152, 4819185715383806976, 4819185650963128320,
 4819180600081639424, 4818980007925865472\\
\hline
NGC~1904: \\2957946042639822336, 2957941124898510336, 2957941021819319808,
 2957940991758294016, 2957940953099859968,\\ 2957940854319622144,
 2957939686088825344, 2957939681789513728, 2957939651729096704\\
\hline
NGC~2808:\\5296796803237729280, 5248765874735697920, 5248765153181145088,
 5248762095164753920, 5248761128785631232,\\ 5248760342817686528,
 5248760235434909696, 5248760102299319296, 5248759823113168896,
 5248759720038803456, \\5248759449464506368, 5248759376441485312,
 5248759135923385344, 5248759037147532288, 5248758964124581888,\\
 5248758895405121536, 5248758624830548992, 5248758590470656000,
 5248756975562514432, 5248756185281133568,\\ 5248756146626162688,
 5248756077906676736, 5248755704252886016\\
\hline
NGC~3201: \\413909669362823168, 5413759104989743104, 5413594418771054592,
 413593456698379264, 5413592421596412928,\\ 5413592181078246400,
 413586309872996352, 5413586309872996352, 5413584317008601088,
 413583767252371456, \\5413582392863237120, 5413581877467150336,
 413581087192722432, 5413579918961600512, 5413578682010973184,\\
 413578063535695872, 5413576890996600832, 5413576616122675200,
 413576586067368960, 5413576169446119424,\\ 5413576135086384128,
 413575658354374656, 5413575654049953792, 5413575589634818048,
 413575486555714560,\\ 5413575486555714560, 5413575447891531776,
 413575383476443136, 5413575177317686272, 5413575074238693376,\\
 413574932495368192, 5413574249605269504, 5413574073501926400,
 413573562410144768, 5413573287532235776,\\ 5413566484303136768,
 413554905071314944, 5413553702480466944, 5413553461962434560,
 413548479800352768,\\ 5413535899826848768, 5413534594158413824,
 413534289227462656, 5413528757309438976, 5413526489566784512,\\
 413525733652639744, 5413525733652537344, 5413524462342242304,
 413519548899830784, 5413517933992117248,\\ 5413496046837951488\\
\hline
NGC~5139:\\ 6084085233675677696, 6083894812006170624, 6083888897833447424,
 083768333843197952, 6083765993045113856,\\ 6083721673317449728,
 083721054842294272, 6083718684019539968, 6083717236590460928,
 083717206550649856,\\ 6083717172190866432, 6083717167861504000,
 083717000392158208, 6083716966032427008, 6083716755544271872,\\
 083716725514207232, 6083716240149382144, 6083715759112312832,
 083715518593904640, 6083715488563689472,\\ 6083715179326546944,
 083714870078915584, 6083714629570586624, 6083714590880916480,
 083714526491330560, \\6083714384722077696, 6083714251613096960,
 083714247283596288, 6083713942363064320, 6083713633137661952,\\
 083713422659409920, 6083713113416041472, 6083713044692316160,
 083712945942896640, 6083712945930018816,\\ 6083712804173831168,
 083712735454282752, 6083712671051794432, 6083712430546763776,
 083712391856739328, \\6083711159236003840, 6083710609480046592,
 083710231522910208, 6083709333841792000, 6083708754054046720,\\
 083708646646635520, 6083708513535848448, 6083708444816310272,
 083708303049198592, 6083708303049198592,\\ 6083708096890249216,
 083708066859057152, 6083707993815319552, 6083707791962719232,
 083707585823037440, \\6083707276585314304, 6083706829908632576,
 083705863503050752, 6083705863503050752, 6083705764756611072,\\
 083705524223446016, 6083705451202263040, 6083705451190582272,
 083705313751263232, 6083705215000783872,\\ 6083705214983351296,
 083704660916108288, 6083704562165564416, 6083704459086503936,
 083704459086503936,\\ 6083704454770907136, 6083704317317423104,
 083704115489007616, 6083704042440567808, 6083703973721079808,\\
 083703565733095424, 6083703497013605376, 6083703325213995008,
 083703084695678976, 6083703084671745024, \\6083703084671715328,
 083703015976199168, 6083703015976195072, 6083702908569564160,
 083702805490460672,\\ 6083702736775139328, 6083702603659175936,
 083702466197170176, 6083702083940176896, 6083701297989063680,\\
 083701297989044224, 6083701126190159872, 6083701091830602752,
 083700950063902720, 6083700881344475136,\\ \hline
\end{tabular}
\end{table}

\setcounter{table}{2}
\begin{table}[h!]
\caption{continued.}
\label{tab3a}
%\small
\begin{tabular}{l}
\hline

6083700851285649408,
 083700816952988672, 6083700782592749568, 6083700473355071488,
 083700469041465344,\\ 6083700061038435328, 6083700061038434304,
 083699957959197696, 6083699884911561728, 6083699854879774720,\\
 083699854879750144, 6083699644393321472, 6083699064605677568,
 083698583569591296, 6083698510527129600,\\ 6083698068152813568,
 083697999433011200, 6083697930735219712, 6083697823327502336,
 083697587137983488,\\ 6083697479729909760, 6083697174799136768,
 083696693784602624, 6083696586374838272, 6083696453266409472,\\
 083696384546921472, 6083696208417679360, 6083695800430514176,
 083695662992438272, 6083695113235747840,\\ 6083694662228403200,
 083693292169089024, 6083692398815815680, 6083689989303946240,
 083689718756243456,\\ 6083676421537754112, 6083675940501016576,
 083675768702706688, 6083670511662609408, 6083518366722814976,\\
 083516408217647104, 6083516403911318528, 6083516266472284160,
 083516064620179456, 6083515441838221312,\\ 6083515441837944832,
 083515373118724096, 6083514724590233600, 6083514685923960832,
 083512143299539968, \\6083511288616147968, 6083511146867133440,
 083510979379388416, 6083509394524092416, 6083509295752195072,\\
 083509257084211200, 6083509227018093568, 6083508879124407296,
 083506615692379136, 6083506439583009792,\\ 6083506306454781952,
 083503660753956864, 6083502939200442368, 6083502831810494464,
 083491359967537152\\
\hline
NGC~5272: \\454892344429695488, 1454880765198055168, 1454880387240926208,
 454880387240926208, 1454880387240926208, \\1454879322088996864,
 454878394376041216, 1454877501022907392, 1454875920474926592,
 454875714316484608,\\ 1454875405078903808, 1454875405078903808,
 454875095841834496, 1454875095841834496, 1454874099408643328,\\
 454874099408643328, 1454874099408643328, 1454873820233046016,
 454873820233046016, 1454873820233046016, \\1454825239860913664,
 454799641858137856, 1454799641858137856, 1454799641858137856,
 454798439267385856,\\ 1454797889511583744, 1454797889511583744,
 454797442834964992, 1454795312531144192, 1454795312531144192,\\
 454795312531144192, 1454785691804450304, 1454785691804450304,
 454784626652863488, 1454784489215173888, \\1454784489214815232,
 454784489214815232, 1454784484916760320, 1454784484916760320,
 454784111256599040,\\ 1454784111256599040, 1454784111256599040,
 454783767659203072, 1454783595860359680, 1454783595860359680,\\
 454783595860359680, 1454781396838245376, 1454781362477636096,
 454781358180631808, 1454780774062209024,\\ 1454780774062209024,
 454780744002202368, 1454777033150420736, 1454777033150420736,
 454776242876414976,\\ 1454776242876414976, 1454775658760780288,
 454770779678034432\\
\hline
NGC~5904: \\421624580189612544, 4421624580189612544, 4421622071928631808,
 421621689671197696, 4421621384733858816,\\ 4421620731898799104,
 421620285222204928, 4421620079063749120, 4421574105733508096,
 421573620401715200,\\ 4421573586041974272, 4421573517322513408,
 421573246740321792, 4421573173724967936, 4421573070645750272,\\
 421572834419569664, 4421572761408122880, 4421572761408105472,
 421572555249780736, 4421572520890057728,\\ 4421571941069952512,
 421571391314576896, 4421571082076932608, 4421571082076932608,
 421570944637486592,\\ 4421526483133363200, 4421525074384092160,
 421518786551972864, 4421518786551972864, 4421479204133344256,\\
 421479135413870592, 4421479066694386688, 4421478929255508992,
 421478684440065024, 4421478684440065024,\\ 4421478516938564096,
 421478169048094720, 4421478169048094720, 4421478169048091136,
 421478134688349184,\\ 4421478100328619520, 4421477692304774144,
 421477344410272768, 4421477279987915776, 4421477142548956160,\\
 421474737367267328, 4421474045875793920, 4421473706575096832,
 421473328617968640, 4421472916301109248\\
\hline
NGC~6121: \\048489472438681600, 6045506806641715200, 6045506806641715200,
 045504190988426240, 6045503679905453056,\\ 6045501961918453760,
 045492135033247744, 6045491482198181888, 6045490382686606336,
 045489180096624640,\\ 6045488767778740224, 6045485709761996800,
 045485366164586496, 6045479490649030656, 6045479078332149760,\\
 045478528576251904, 6045478356777482240, 6045478047540140032,
 045477807021629440, 6045477807021629440, \\6045477356033738752,
 045476363912671232, 6045475711077568512, 6045475333120344064,
 045473374615218176,\\ 6045466640106128384, 6045466360918967296,
 045466330868396032, 6045466326563297280, 6045465987271055360,\\
 045465952911275008, 6045465746752802816, 6045465746752765952,
 045465643673558016, 6045465403155228672,\\ 6045465368795453440,
 045465265716182016, 6045465231356507136, 6045465093917682688,
 045465055252946944,\\ 6045464956478708736, 6045464887759156224,
 045464814730497024, 6045464578521393152, 6045464269284031488,\\
 045463856967250944, 6045463444650191872, 6045462860534542336,
 045462688735810560, 6045462276418864128, \\6045462169030878208,
 045461967181168640, 6045461172595996672, 6045460524072098816,
 045460420992873472,\\ 6045460214834413568, 6045459939956596736,
 045457328616272896, 6045456778860434432, 6045409255034381312,\\
 045394656451832832\\
\hline
NGC~6273:\\ 4112105459935354368, 4112103260912084480, 4111917374764549120,
 4111916034729051648, 4111914930863089664, \\4111914900857380864,
 4111914518546191360\\
\hline
NGC~6397:\\ 921759002005352448, 5921754058520058880, 5921753504449513472,
 921751649029137408, 5921751034863718400,\\ 5921748762804733952,
 921748247414961152, 5921747869451541504, 5921744815749877760,
 921744296038758400, \\5921744089880325120, 5921743888036909056,
 921704649213888512\\
\hline
NGC~6656: \\077591652549154304, 4077589178647903232, 4077589036825337856,
 077588800710484992, 4077588349639098880,\\ 4077587426300886016,
 077494792457764352, 4077494719343533056, 4077494410105892864,
 077493963429298176,\\ 4077493585472178176, 4077485476573937664,
 076837314472061440, 4076821783873141760, 4076742069274543616,\\
 076741798780045312, 4076741004122655744, 4076738049185903104,
 076736296841467904, 4076733930399045120\\
\hline
\end{tabular}
\end{table}

\setcounter{table}{2}
\begin{table}[h!]
\caption{continued.}
\label{tab3a}
%\small
\begin{tabular}{l}
\hline

NGC~6715: \\6762054961052211200, 6762051387639623680, 6761953737270846464,
 761842583509689344, 6761669857110812672,\\ 6761619451372360704,
 761602920021280768, 6761404084548322304, 6761404084548322304,
 761398037230049280,\\ 6761350792588121088, 6761340415946236928,
 761339346470157312, 6761337804606019584, 6761336425892378624,\\
 761330520341490688, 6761326461568118784, 6761318631871929344,
 761318631871929344, 6761247434169513984,\\ 6761181983190000640,
 761180810634108928, 6761180535756236800, 6761173049658036224,
 761173049658036224,\\ 6761173049658036224, 6761173049658036224,
 761080931200651264, 6761080931200651264, 6760819041266382848,\\
 760542303626063872, 6760478536255302656, 6760444829365898240,
 760444829365898240, 6760435930193634304,\\ 6760434418365123584,
 760434418365123584, 6760433898643507200, 6760431463427546112,
 760431463427546112,\\ 6760431463427546112, 6760425892824511488,
 760424831996624896, 6760423762520598528, 6760416237737643008,\\
 760397619083550720, 6760397619083550720, 6760397619083550720,
 760396794449819648, 6760394251828638720,\\ 6760394251828638720,
 760394251828638720, 6760386486527685632, 6760384012604475392,
 760374632417670144,\\ 6760374632417670144, 6760374632417670144,
 760373601625835520, 6760373601625835520, 6760360063887816704,\\
 760353531214773248, 6760353531214773248, 6760325944667185152,
 760325944667185152, 6760324398478152704,\\ 6760324398478152704,
 760324398478152704, 6760324398478152704, 6760320790705521664,
 760320790705521664,\\ 6760320790705521664, 6760253342536820736,
 760253342536820736, 6760235372393313280, 6760235372393313280,\\
 760235372393313280, 6760231386663496704, 6760231386663496704,
 760231386663496704, 6760231386663496704, \\6760178781902570496,
 760178781902570496, 6760177648031284224, 6760177648031284224,
 760177648031284224,\\ 6760176956525425664, 6760176956525425664,
 760176720318282752, 6760176720318282752, 6760170909211427840,\\
 760170909211427840, 6760167374469435392, 6760163423099388928,
 760163423099388928, 6760159536131468288, \\6760157581943897088,
 760135381242258432, 6760135381242258432, 6760135381242258432,
 760134900198578176,\\ 6760128474927442944, 6760127100538032128,
 760127036135541760, 6760126108422582272, 6760123462722973696,\\
 760071235918557184, 6760046943583289344, 6757470272440285184,
 757050465157275648, 6757044280404147200,\\ 6757031567300829184,
 737310211366472704, 6737076251582371840\\
\hline
NGC~6752: \\638394262652962816, 6638394022134790144, 6638381201651214336,
 638380686257568768, 6638378079215826944,\\ 6638377976136644608,
 638377495100327936, 6638377250280690688, 6638376258149705728,
 638331723627352064,\\ 6638319525920241664, 6632389589136922624,
 632388725840159744, 6632388489625249792, 6632386977796837376,\\
 632384091578990592, 6632376631212259328, 6632376429357002752,
 632375806581891072, 6632375669142896640, \\6632375256822736896,
 632372576766505984, 6632371893871384576, 6632371034877872128\\
\hline
NGC~6809: \\6751396089162510336, 6751391961690798080, 6751391759835333632,
 6751389075477279744, 6751388594440936448,\\ 6751343411385100288,
 6751295891862253568\\
\hline
NGC~6838: \\
821627988289201664, 1821621116340655616, 1821621081980856576,
821620841462681088, 1821620738383535104,\\ 1821620497865339392,
821620291706888704, 1821619913749895680, 1821613690298017536,
821608952991943168,\\ 1821608712484571392, 1821608712473895424,
821608678122974720, 1821608575034967808, 1821607990919194880,\\
821607990919181824, 1821607956559392768, 1821607922199682048,
821607269364578560, 1821606783991599104\\
\hline
\end{tabular}
\end{table}

\twocolumn
\end{appendix}

\end{document}